\documentclass[journal]{IEEEtran}

\usepackage{cite}
\usepackage[fleqn]{amsmath}
\usepackage{latexsym}
\usepackage{amsfonts}
\usepackage{accents}
\usepackage{cite}
\usepackage{url}
\usepackage{algorithm}
\usepackage{algorithmic}
\usepackage{times}
\usepackage{epsfig}
\usepackage{graphicx}
\usepackage{amssymb}
\usepackage{mathtools}
\usepackage[abs]{overpic}
\usepackage{multirow}
\graphicspath{{./images/}} 

\def\bq{\begin{equation}}
\def\eq{\end{equation}}
\def\beq{\begin{eqnarray}}
\def\eeq{\end{eqnarray}}
\def\ba{\begin{array}}
\def\ea{\end{array}}
\def\bc{\begin{center}}
\def\ec{\end{center}}

\def\e-jwi{e^{-j\omega_{i}}}

\DeclareSymbolFont{lettersA}{U}{txmia}{m}{it}
 \DeclareMathSymbol{\FF}{\mathord}{lettersA}{'206}
 \DeclareMathSymbol{\NN}{\mathord}{lettersA}{'216} 
 \DeclareMathSymbol{\RR}{\mathord}{lettersA}{'222} 
 \DeclareMathSymbol{\ZZ}{\mathord}{lettersA}{'232} 
 \DeclareMathSymbol{\QQ}{\mathord}{lettersA}{'221} 
 \DeclareMathSymbol{\CC}{\mathord}{lettersA}{'203}

\def\b0{\mbox{\boldmath$0$}}
\def\b1{\mbox{\boldmath$1$}}

\def\mb1{\mathbf{1}}

\def\mbn{\mathbf{n}}

\def\mbs{\mathbf{s}}

\def\mbx{\mathbf{x}}
\def\mby{\mathbf{y}}

\def\mbA{\mathbf{A}}

\newcommand{\bhline}[1]{\noalign{\hrule height #1}}
\begin{document}
\title{Self-supervised Hyperspectral Image Restoration using Separable Image Prior}

\author{Ryuji~Imamura,~\IEEEmembership{NonMember,}
Tatsuki~Itasaka,~\IEEEmembership{NonMember,}
        Masahiro~Okuda,~\IEEEmembership{Senior Member, ~IEEE,}% <-this % stops a space
\thanks{This work was partially supported by the Ministry of Internal Affairs and Communications of Japan under the Strategic Information and Communications R\&D Promotion Programme of 2017, No. 3620.}% <-this % stops a space
\thanks{The authors are with the Department of Information System Engineering, The University of Kitakyushu, Japan (email: okuda-m@kitakyu-u.ac.jp). }% <-this % stops a space
}

\markboth{Journal of \LaTeX\ Class Files,~Vol.~14, No.~8, August~2019}%
{Shell \MakeLowercase{\textit{et al.}}: Bare Demo of IEEEtran.cls for IEEE Journals}

\maketitle

\begin{abstract}
Supervised learning with a convolutional neural network is recognized as a powerful means of image restoration. However, most such methods have been designed for application to grayscale and/or color images; therefore, they have limited success when applied to hyperspectral image restoration. This is partially owing to large datasets being difficult to collect, and also the heavy computational load associated with the restoration of an image with many spectral bands. To address this difficulty, we propose a novel self-supervised learning strategy for application to hyperspectral image restoration. Our method automatically creates a training dataset from a single degraded image and trains a denoising network without any clear images. Another notable feature of our method is the use of a separable convolutional layer. We undertake experiments to prove that the use of a separable network allows us to acquire the prior of a hyperspectral image and to realize efficient restoration. We demonstrate the validity of our method through extensive experiments and show that our method has better characteristics than those that are currently regarded as state-of-the-art.
\end{abstract}


\IEEEpeerreviewmaketitle

\section{Introduction}
\IEEEPARstart{H}{yperspectral} image (HSI) restoration, whereby a degraded image is processed to produce a clear image, is an essential technique in HSI applications \cite{introHSI,dspHSI2,mia1,dspHSI1} given that hyperspectral images are prone to being adversely affected by noise, which prevents the precise extraction of useful information in applications such as classification, unmixing, and target detection \cite{introUnmix,mia2,okabe,introTarget,introClass}. 
Various HSI restoration methods have been proposed \cite{STV, ASTV, MTTV, SSAHTV, CTV}. Optimization-based methods with image priors have constituted one of the mainstreams in the HSI reconstruction, and have been the subject of considerable research. Some methods have exploited the sparsity, low-rank properties, and non-local correlation of HSIs to restore images by solving optimization problems based on the priors \cite{SSTV,LRTV,LRTDTV,SSAHTV,LRMR}. 

The hyperspectral total variation proposed in \cite{SSAHTV} is an expansion of the classical total variation (TV) model to an HSI.
 A structure tensor TV (STV) \cite{STV} was proposed for multi-band images. This considers low-rankness of gradient vectors in local regions. Furthermore, some modified versions were proposed in \cite{ASTV,eusipco17}, which was capable of outperforming STV in some image-restoration problems. Furthermore, a spatio-spectral TV (SSTV) was proposed in \cite{SSTV}; this applies a TV to the HSI gradient in the spectral direction. As such, it constitutes a regularization scheme that indirectly considers the spatial correlation through the application of TV to the spectral gradient. 
 
It is known that the spectral information of an HSI is well represented by a few pure spectral endmembers. The property induces the low-rankness of an HSI. There are many methods that exploit this low-rankness.
 Low-rank matrix recovery (LRMR) \cite{LRMR}, which utilizes the low-rankness in a local area of the HSI, has been shown to offer a high level of performance. 
Unlike STV and ASTV, noise tends to remain in the restoration results of LRMR and SSTV because of the weak influence of the spatial properties. Other methods that exploit the low-rankness, such as LRTV \cite{LRTV} and LRTDTV \cite{LRTDTV}, attempt to improve the performance of STV, SSTV, and its variants \cite{HSSTV} by newly introducing regularization. These methods offer excellent performance, especially when a model matches a real degradation process, and offer an advantage in that learning with a large dataset is not necessary. However, future tasks, such as over-smoothing, have yet to be elucidated.
Most of these methods can be regarded as convex optimization problems and can produce clear images as global optimums. 

Exploiting the self-similarity of images plays a crucial role in some image restoration methods. BM4D \cite{BM4D}, which is built based on the well-known non-local method BM3D, has been used as a baseline to verify the performances of HSI restoration methods. FastHyDe \cite{FastHyDe} efficiently exploits both of the low-rankness and self-similarity. Although these methods realize high performance, they are suitable only for Gaussian and Poisson noises.

Aside from the methods that rely on designed priors, deep learning has been successfully applied to various computer vision and image processing tasks, while image restoration based on deep neural networks (DNNs) has recently been the subject of research. It has been shown that restoration with a high degree of precision can be attained \cite{MLP,TNRD,CSF,DnCNN} through the use of large supervised datasets. Although deep learning becomes integrated into image-restoration tasks, the high performance is limited to the applications of grayscale/color images. 
Regarding the HSIs, although a few methods have been proposed for HSI restoration, such as \cite{hsi_deep2, hsi_deep1, hsi_deep3}, DNN is only slightly superior for HSI restoration. This is partly owing to the difficulty of acquiring large amounts of data as a result of the need for special sensors to capture the HSIs, as well as the time and effort needed to obtain images.
%This is a result of it being extremely difficult to collect large datasets required for HSIs for supervised training, because special HSI sensors are required, and it often takes much time and effort to capture an HSI. 

\begin{figure}[t]
\centering
\includegraphics[width=0.9\linewidth]{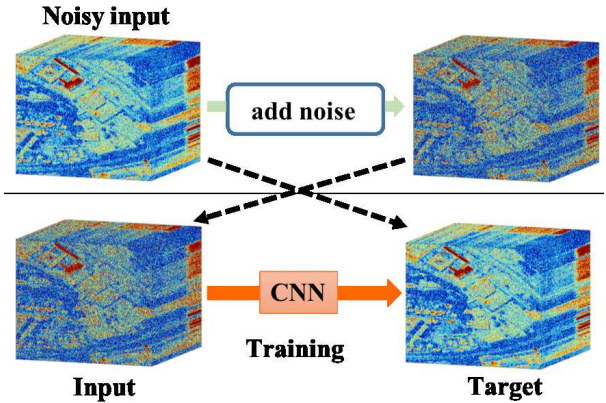}
\caption{Self-supervised restoration: (top) We add noise to an noisy input. (bottom) We train CNN by setting the input as a target and the image with the added noise as an input for CNN. }
\label{fig:abst}
\end{figure}

Another problem is that, to fully exploit the features of an HSI through the application of deep learning, a large-scale DNN is required. An HSI typically has more than a hundred spectral bands and, thus, to exploit the correlation between an HSI in the spectral domain, we need intermediate feature maps with many more channels in the DNN. 
For example, DnCNN \cite{DnCNN} uses 17--20 layers with 64 filter kernels in each layer to train the denoising network to handle color images. Therefore, based on the DnCNN settings, it would be preferable to use $64 \times N/3$ filters in each layer to capture the spectral information of an HSI with $N$ spectral bands. Therefore, any increase in the number of parameters makes it difficult to train a network.

To circumvent these difficulties, we propose a novel DNN-based learning scheme for HSI restoration by introducing a new self-supervising strategy. Our method creates input/target patches from a single deteriorated image and trains denoising networks without any clear images (Figure \ref{fig:abst}). Another feature of our method is that, considering the anisotropic property of HSIs, we can utilize a separable convolutional network. We show that the proposed network with the separable convolution efficiently captures an HSI prior, through experiments involving some image-restoration tasks. To the best of our knowledge, this is the first attempt to apply self-supervised learning to HSI restoration with DNNs. The results of our experiments showed that our method is comparable to or outperforms conventional methods based on model-based methods.

%
%%%%%%%%%%%%%%%%%%%%%%%%%%%%%%%%%%%%%%%%%%
\section{Separable CNN Capturing Structural Prior}
\label{sec:label}
%%%%%%%%%%%%%%%%%%%%%%%%%%%%%%%%%%%%%%%%%%
\subsection{Conventional self-learning approaches}
\label{sec:conv}
%%%%%%%%%%%%%%%%%%%%%%%%%%%%%%%%%%%%%%%%%%
A CNN, composed of linear filters, is inherently able to represent a signal with a high correlation. Ulyanov et al. \cite{DIP} exploited this property using Deep Image prior (DIP), proposing a generative network that maps a random noise vector to a natural image. They demonstrated that the structure of the CNN itself encloses the statistical prior of the images and exhibits excellent performance with some restoration tasks, such as denoising, super-resolution, and inpainting. Lehtinen et al. \cite{N2N} used a noisy image pair to learn the "{\it noise to noise}" relationship between the two measurements, showing that it is possible to produce a clear image using the smoothing property of the CNNs. They experimentally demonstrated that a clear image, which is often hard to obtain, is not needed to train a denoising network. Krull et al. \cite{N2V} built on this idea to develop self-supervised learning and proposed a method called Noise2Void (N2V), in which an input image is predicted using a convolutional kernel without a center coefficient. DIP and N2V do not require us to make any assumptions for the degradation process of an input image. However, the use of DIP often gives rise to over-smoothing, especially when applied to the restoration of severely degraded images, while the performance of N2V becomes worsen when an image does not have a high correlation in a local region, such as images with many holes.
Shocher et al. \cite{ZeroshotSR} proposed another approach to self-supervised learning for image super-resolution.  They assumed that the statistical properties of an image do not change with the scale, and that the relationship between an original image and its downsized version is learned using a smaller image pair that is made based on the measurements.

Our work was inspired by these methods. In this section, we propose a novel self-supervised strategy and introduce a separable convolutional network for HSI restoration. We show that the separable structure is highly suited to the HSI restoration tasks, achieving high-fidelity restoration by our self-supervised learning.
\subsection{Low-rank property of Hyperspectral images}
\label{sec:lowrank}
%%%%%%%%%%%%%%%%%%%%%%%%%%%%%%%%%%%%%%%%%%
%
%
\begin{figure}[t]
\centering
\includegraphics[width=0.9\linewidth]{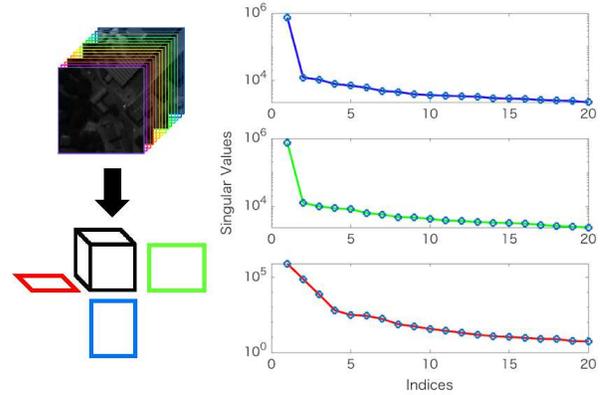}
\caption{(left) Tucker Decomposition of HSI, (right) Plot of singular values for each mode (shown in log scale).}
\label{fig:tucker}
\end{figure}

 It is well known that a measured spectrum in a hyperspectral pixel can be approximated by a small fraction of spectral signatures, called endmembers. In an aerial scene, endmembers correspond to familiar macroscopic materials. In general, endmembers in a neighbor tends to be highly correlated, which leads to the low-rankness of the HSI, especially in the spectral direction.

An HSI can be viewed as a third-order tensor, the two dimensions of which represent the spatial characteristics, while the third dimension corresponds to spectral bands. Figure \ref{fig:tucker} illustrates the Tucker decomposition of an HSI with singular value plots for all three modes. We can see that all the plots exhibit rapid decaying trends, pointing to the low-rankness of the image. Comparing the three plots in Figure \ref{fig:tucker}, the low-rankness in the third mode (indicate by the red plots) is more remarkable, implying that the $y-z$ (or $z-x$) planes (i.e. planes including the spectral direction) in the tensor can be approximated by smaller dimensions than the other planes (green and blue plots).

\begin{figure}[t]
\centering
\includegraphics[width=0.9\linewidth]{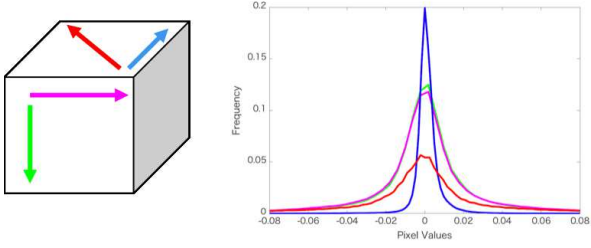}
\caption{Four plots indicate histogram of difference of adjacent pixels: (magenta) horizontal direction, (green) vertical direction, (blue) spectrum direction, (red) diagonal direction in spatio-spectral plane }
\label{fig:correlation}
\end{figure}

In Figure \ref{fig:correlation}, the four plots show the histograms of the difference of adjacent pixels with respect to $x, y, z$ and the diagonal directions of the $y-z$ plane. To obtain the plots, we applied a differentiable filter to planes along with each direction and calculated the histograms. A high peak around zero indicates a high correlation. It is clear that the diagonal correlation (indicated by the red line) is less than that in the other three directions. These properties hold for most HSIs. These directional dependencies allow us to separately treat an HSI in the spatial and spectral directions.

In the following section, to fully exploit this a priori knowledge, we introduce CNN-based image reconstruction using separable convolution. We experimentally demonstrate that the separable CNN can capture the properties of a HSI, making it suitable for many HSI restoration tasks.

%%%%%%%%%%%%%%%%%%%%%%%%%%%%%%%%%%%%%%%%%%
%%%%%%%%%%%%%%%%%%%%%%%%%%%%%%%%%%%%%%%%%%
\subsection{Architecture}
%%%%%%%%%%%%%%%%%%%%%%%%%%%%%%%%%%%%%%%%%%
%%%%%%%%%%%%%%%%%%%%%%%%%%%%%%%%%%%%%%%%%%
%%%%%%%%%%%%%%%
\begin{figure*}[!t]
  \centering
\includegraphics[width=0.8\linewidth]{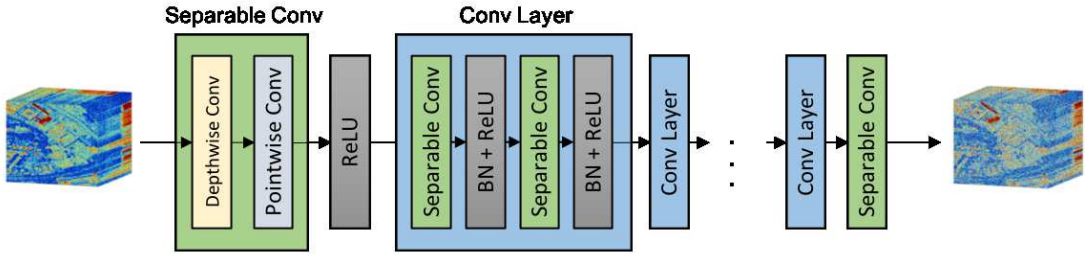}
\caption{Architecture: We use a simple separable CNN composed of depth-wise and point-wise convolutional layers. We use $M$ kernels for a single depth-wise layer, and its $M$ outputs are concatenated with respect to the channel direction. }
\label{fig:arch}
\end{figure*}

A non-separable CNN with dense coefficients is inherently better able to capture the features of images than a separable CNN with the same number of layers, but it is often hard to train it for an image cube with many channels like a HSI, because we may need more kernels for each layer to capture the features in the spectral direction. This would make its optimization very difficult. Considering the property of HSIs as discussed in Section \ref{sec:lowrank}, we can safely ignore diagonal correlations in the spatial-spectral plains because less information is carried by the neighboring pixels in the diagonal direction, pointing to the advantage of using a separable network. 
In this section, we introduce the architecture with the separable CNN and examine, through self-supervised image restoration, that the separable architecture not only reduces the training complexity but also encloses the structural advantage to represent an HSI.

Figure \ref{fig:arch} illustrates our network that includes separable convolutional layers, batch normalization units, and ReLU. 
Separable convolution \cite{Xception} was originally introduced to reduce the computational complexity of CNN by approximating 3D convolution to separate convolutions in the spatial and channel dimensions.
Convolution in the spatial dimension is performed independently in each channel (depth-wise convolution), while convolution in the channel direction is performed using a 1D kernel (point-wise convolution). 

Let H, W be the dimensions of an input image, and M, L be the channel lengths of the input and output of a separable convolution, respectively.
In our architecture, we use $N$ 2D kernels for a single depth-wise layer. A set of 2D kernels is represented by a 3D array ${\bf W}_d\in \mathbb{R}^{H\times W\times N}$. The outputs of all the kernels in ${\bf W}_d$ are concatenated with respect to the channel direction, resulting in an output ${\bf h_{\it d}}\in\mathbb{R}^{H\times W\times M\cdot N}$ of the depth-wise convolution, given by 

\begin{equation}%depth wise convolution
\begin{split}
&{\bf h_{\it d}}(h,w,m')=\sum^{K}_{k_1=1}\sum^{K}_{k_2=1}{\bf W}_d(k_1,k_2,m){\bf h_{\it in}}(h',w',n),\\
&\quad(h' \coloneqq h+\tfrac{K+1}{2}-k_1, w'\coloneqq w+\tfrac{K+1}{2}-k_2,\\
& \quad\qquad m'=m\cdot M +n )
%& i=1,2,\cdots, H; j=1,2,\cdots, W; \quad m= j=1,2,\cdots, D_{in} )
\end{split}
\label{eq:depth_conv}
\end{equation}
where the kernel size is $(K \times K)$, and the parenthesized indices represent the position of an element in this variable. Similarly, the output of the point-wise convolution with 1D kernels is represented by a 2D variable ${\bf W_{\it p}}\in \mathbb{R}^{M\cdot N \times L}$
\begin{equation}
\label{eq:point_conv}
{{\bf h_{\it p}}}(h,w,l)=\sum^{M\cdot N}_{m'=0}{\bf W_{\it p}}(l,m')\cdot{\bf h_{\it in}}(h,w,m').
\end{equation}
In the next section, we evaluate the validity to use this separable convolution for the HSI restoration.

%
%
%%%%%%%%%%%%%%%
\subsection{Performance of Separable CNN}

We now demonstrate that a network with a separable convolution can efficiently capture the structural prior of the HSIs by considering a simple example of image hole-filling. 
Let the linear operator $\mbA \in \{0,\,1\}^{N\times N}$ denote a known random-sampling matrix, where $N$ is the total number of pixels in an HSI. Given only the pixels corresponding to the sampling points of $\mbA$, we consider the problem of recovering an entire image $\mbx$. 

Here, we train a network by minimizing the mean squared error only for those pixels masked by $\mbA$, that is, 
\begin{equation}
\min_{\Phi} \| \mbA\mbx-\mbA\Phi(\mbA\mbx) \|_2^2,
\label{eq:cs}
\end{equation}
where $\mbA\mbx$ is a set of measured pixels, and we denote a neural network by $\Phi$. The aim of the task is to estimate an original image $\mbx$.

We trained two networks by minimizing \eqref{eq:cs}. One of these was a non-separable CNN with tightly coupled coefficients. It had eight layers, with each layer having $3\times 3\times M \times L$ parameters. The other was a separable CNN like that shown in Figure \ref{fig:arch}. This also has eight layers, with each layer having $3\times 3\times 1 \times M$ parameters for the depth-wise convolution and $1\times 1\times M \times L$ parameters for the point-wise convolution. In the experiment, we divided an HSI into $20 \times 20\times \text{\{number of spectral bands\}}$ blocks and used them as inputs. The training was carried out by carefully tuning the hyper-parameters to attain the best performance. Note that this training with \eqref{eq:cs} differs from DIP in that we used the image as is, while DIP starts with random samples to generate a clear image. 

\begin{figure}[t]
\centering
\includegraphics[width=0.45\linewidth]{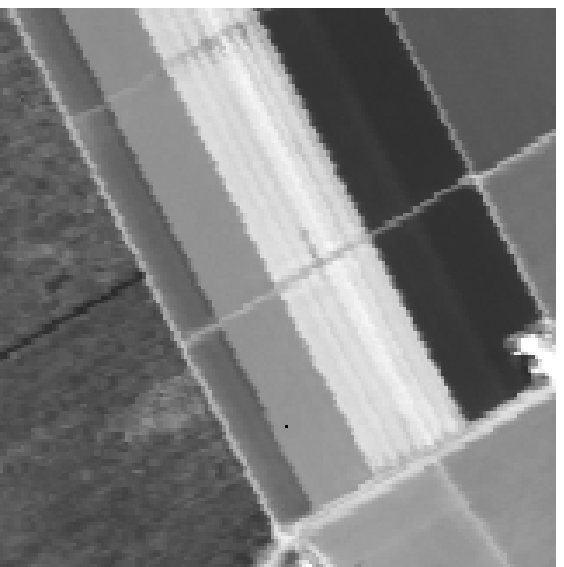}
\includegraphics[width=0.45\linewidth]{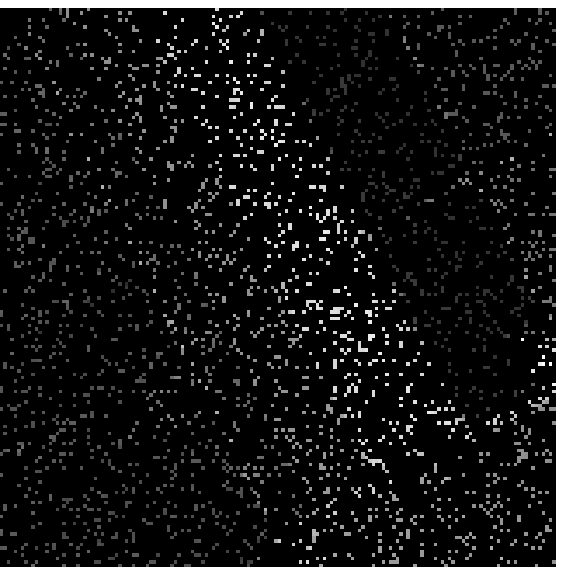}\\
(a) \hspace{3.5cm}(b)\\
\includegraphics[width=0.45\linewidth]{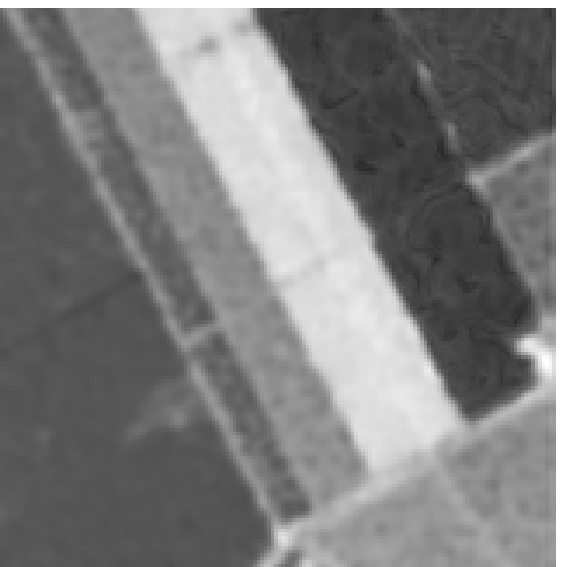}
\includegraphics[width=0.45\linewidth]{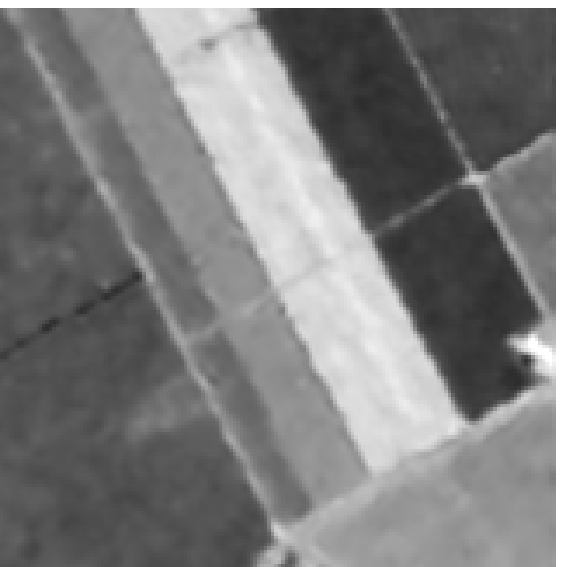}\\
(c) PSNR: 37.88 \hspace{1.5cm} (d) PSNR: 39.13
\caption{Self-supervised hole-filling: (a) original image, (b) observation, (c) result obtained with dense network, (d) results obtained with separable network (PSNR values indicate the mean of PSNR for all bands)}
\label{fig:cs}
\end{figure}

Figure \ref{fig:cs} illustrates an example of the hole-filling task. Figure \ref{fig:cs} (a) and (b) show the original image and an observation, respectively.  The results obtained with the non-separable and separable CNN are shown in Figure \ref{fig:cs} (c) and (d), respectively. Both of the networks interpolates the holes even though the cost is evaluated only on the sampling points indicated by $\mbA$ by virtue of smoothing capability of the CNNs. It should be noted that even though the separable CNN has the same number of layers and fewer parameters, the performance is superior to that of the non-separable CNN. 

%%%
%Remarks
%%%
The CNN with linear filters is inherently able to represent natural images, as demonstrated in \cite{DIP}.
Considering the task of estimating a clear image $\mbx$ from a degraded observation $\mby$, the optimization problem can be expressed as described in \cite{DIP}
% $\max_{\mbx} \log p(\mby|\mbx) + \log  p(\mbx)$, which 
\begin{equation}
\min_{\mbx} \mathcal{F}(\mbx, \mby) + \iota_{S_{\Phi}}(\mbx),
\end{equation}
where $\mathcal{F}(\mbx, \mby) $ is a fidelity term that is usually determined by a degradation model. $S_{\Phi}$ indicates a set of all the images that can be represented by a neural network $\Phi$. The prior  $\iota_{S_{\Phi}}(\mbx)$ is an indicator function given by
\begin{equation}
\iota_{S_{\Phi}}(\mbx) =\left\{ 
\begin{array}{cc} 0,& \text{if } \:\mbx \in S_{\Phi}\\ +\infty,& \text{otherwise}\end{array}
\right.
\end{equation}

Roughly speaking, if the set $S_{\Phi}$ is small, the prior works well for restricting solutions. 
Evidently, the set of images that can be represented by a separable network is a subset of those that can be represented by a dense network. Thus, an optimal solution with the dense network is always better than or equal to one with the separable network. 
%Our aim is to add some restriction on a structure of a network such that the prior well represents a priori knowledges of HSIs.

Through experiments, however, we found that we could optimize the separable CNN quite easily, while the non-separable CNN was barely able to fit the features of an HSI, making its restoration capability relatively low. This superiority of the separable CNN was confirmed not only with this hole-filling task but also with other restoration tasks. Our experiments showed that the separable network achieves a better solution than the dense network due to the ease of training.

Obviously, identical mapping incurs zero cost in \eqref{eq:cs}, making it an optimal solution. However, we confirmed through experiments that, when training with random initial weights, it rarely converges to trivial identical mapping, and is instead inclined toward the restoration of an image in most cases.

%%% figure for compressed sensing

%%%%%%%%%%%%%%%%%%%%%%%%%%%%%%%%%%%%%%%%%%
%%%%%%%%%%%%%%%%%%%%%%%%%%%%%%%%%%%%%%%%%%
\section{Self-supervised Training}
\subsection{Gaussian Denoising}
\label{sec:Gauss}
%%%%%%%%%%%%%%%%%%%%%%%%%%%%%%%%%%%%%%%%%%
%%%%%%%%%%%%%%%%%%%%%%%%%%%%%%%%%%%%%%%%%%
%%%%%%%%%%%%%%%%%%%%
%%%%%%%%%% Table Gaussian
%%%%%%%%%%%%%%%%%%%%
\begin{table*}[t]
\caption{Results for Gaussian noise removal}
\setlength{\tabcolsep}{3mm}
\begin{center}
\begin{tabular}{cccccccccc} \bhline{1pt}
Methods & $\sigma$ & PaviaC & PaviaU & Frisco & Stanford & {\footnotesize IndianPine} &  {\footnotesize Washington} & Cupurite \\ \hline \hline
BM4D \cite{BM4D} &0.05& 38.54&	38.53&	40.41&	40.74&	38.01&	39.59 &	38.14 \\
 &0.1& 34.94&	34.94&	36.74&	37.04&	35.26&	35.58 	&34.77 	 \\
  &0.15& 32.83&	32.71&	34.75	&35.04	&33.69	&33.41 	&32.99 	 \\
     &0.2&31.4&	31.15&	33.39&	33.67&	32.55&	31.98 &	31.83   \\
    &0.25& 30.35&	30.00&	32.33&	32.65&	31.65&	30.89 &	31.00  \\
 FastHyDe \cite{FastHyDe}
 &0.05& {\bf 40.68}&	{\bf 40.23}& {\bf 45.78}& {\bf 46.08}&  38.96&  {\bf 44.86} & 39.57  \\
 &0.1& 37.56&	37.27&	41.78&	42.19&	37.29&		40.71	&37.33   \\
 &0.15&{\bf 35.76}&	35.35&	39.57&	40.20&	36.04&		38.38 	&36.18   \\
 &0.2&34.39&	34.11&	38.20&	38.76&	35.17&	36.80 &	35.39   \\
 &0.25&33.46&	33.05&	37.13&	37.39&	34.48&	35.71 &	34.81   \\ \hline
DIP \cite{DIP}
&0.05&34.87& 	34.45& 	37.72 &	37.18 &	36.12& 	33.84& 	37.23     \\
&0.1&32.74& 	32.85& 	33.82 &	32.43 &	31.65& 	30.56& 	36.90     \\
&0.15&30.13& 	30.52& 	31.08 &	28.70 &	28.42& 	27.23& 	35.60    \\
&0.2&27.02& 	27.91& 	27.61 &	25.65 &	25.99& 	24.71& 	33.79   \\
&0.25&25.40 &	25.62 &	25.45 &	24.08 &	24.12& 	22.63& 	32.12  \\
N2V \cite{N2V}
&0.05&32.32 &	31.32 &	35.57 &	37.20 &	33.42 &	35.30& 	33.68  \\
&0.1&31.77 &	30.64 &	35.35 &	36.14 &	34.00 &	34.57& 	33.03   \\
&0.15&31.38 &	30.36 &	35.26 &	35.57 &	33.81 &	33.78& 	32.72   \\
&0.2&31.04 &	30.10 &	34.56 &	34.64 &	33.34 &	33.04& 	32.32   \\
&0.25&30.73& 	29.71 &	34.06 &	34.11 &	32.90 &	32.73& 	32.06    \\ \hline
Ours 
&0.05     & 38.84 &    	38.66 &	42.62 &	        45.78 &	       {\bf 39.74} & 44.09& 		{\bf 40.32 }\\
&0.1&{\bf 37.60}& 	{\bf 37.65} &	{\bf 41.81 }&	{\bf 42.29} &	{\bf  37.66} & {\bf 41.43}& 		{\bf 38.94 }  \\
&0.15&35.73 &	{\bf 35.87} &	{\bf 39.76 }&	{\bf 40.35 }&	{\bf 36.48 }&  {\bf 	39.52} &		{\bf 37.51}  \\
&0.2&{\bf 34.87} &	{\bf 34.81 }&	{\bf 38.72 }&	{\bf 39.02 }&	{\bf 35.55 }&  {\bf 	37.85}& 		{\bf 36.72 } \\
&0.25&{\bf 33.96 }&	{\bf 33.73} &	{\bf 37.58}& 	{\bf 37.90 }&	{\bf 34.75}&	 {\bf 36.63}& 		{\bf 35.86 }  \\ \bhline{1pt}
\end{tabular}
\end{center}
\label{tab:Gauss}
\end{table*}

We consider a degradation model with additive noise:
\begin{equation}
\mby =\mbx +\mbn, 
\end{equation}
where $\mbn$ denotes the Gaussian noise with a standard deviation $\sigma$. Our goal is to estimate a latent clear image $\mbx$, given a degraded observation $\mby$.

Our denoising method builds upon the fact that an HSI has rich information even when it is degraded by noise, and that the separable convolution captures the latent low-rank structure of an HSI. 
Our method involves finding a separable CNN that can remove noise by learning only with a degraded input. In the training step, we automatically create a target image and minimize the error between the input and the target (Figure \ref{fig:abst}).

We assume that a network that can recover a clear image $\mbx$ from an observation $\mby$ also has the ability to recover $\mby$ from $\mby+\mbn$. Based on this assumption, we first estimate the standard deviation $\sigma'$ of the noise from the noisy image $\mby$ and further add noise with $\tilde{\sigma}=(1+\alpha)\sigma',\,(\alpha << 1)$ to $\mby$, as follows:
\begin{equation}
\tilde{\mby} = \mby + \tilde{\mbn},
\end{equation}
where $ \tilde{\mbn} \sim \mathcal{N}(0,\,\tilde{\sigma}^{2})$.
We adopted the method proposed in \cite{estSigma} to estimate the standard deviation. Considering the estimation error, we randomly sample the value of $\alpha$ within a range of $[-0.1,\,0.1]$ and created training sets
$\{\mby' , \mby\}$ during the training steps.
We optimize the network with the training sets $\{\mby' , \mby\}$ by minimizing a simple loss function with a $\ell_2$ norm:
\begin{equation}
\min_{\Phi} \| \Phi(\tilde{\mby}) - \mby\|_2^2,
\end{equation}
where $\Phi$ is the separable CNN shown in Figure \ref{fig:arch}.
This strategy efficiently trains the network and yields sufficient performance, but we further enhance the performance by replacing the input image $\mby$ with a restored image every a few hundred epochs.
%%%%%%%%
\subsection{Mixed anomalies}
\label{sec:Mix}
%%%%%%%%
Next, we considered the task of mixed anomaly removal, where we assume that an image deteriorates according to the model:
\begin{equation}
\mby =\mbx +\mbn+\mbs.
\label{eqn:mix}
\end{equation}
where $\mbn$ again denotes the Gaussian noise, and $\mbs$ denotes other anomalies that appear, albeit sparsely, in an image.

To tackle this restoration task, we use two separable networks, $\Phi_1$ and $\Phi_2$, and find them by minimizing the following loss function,
\begin{equation}
\min_{\Phi_1, \Phi_2} \| \Phi_1(\Phi_2(y)+n')- \Phi_2(y) \|_2^2+ \lambda \| y- \Phi_2(y)\|_1.
\label{eqn:loss_mix}
\end{equation}
%This is in part inspired by the robust principal component decomposition \cite{RPCA}.
This loss function is designed to remove the Gaussian noise $\mbn$ and the sparse anomalies $\mbs$ by $\Phi_1$ and $\Phi_2$, respectively. The second term of \eqref{eqn:loss_mix} evaluated by $\ell_1$ norm  plays a role of sparse noise removal, while the first term is based on the idea of the self-supervised strategy introduced in Section \ref{sec:Gauss}. 

For this task, we add various levels of Gaussian noise to a noisy input instead of estimating the standard deviation. 
The trained function $\Phi_1$ finally yields a restored image by $\mbx^*=\Phi_1(\mby)$.

%%%%%
\subsubsection*{Differences from other types of self-supervised restoration}
The proposed method is inspired by DIP and N2V in that both the methods use a degraded measurement as a target; however, unlike those methods, our approach uses a further-degraded image. DIP leverages the CNN property whereby a smoother and clearer image appears first, with noise being generated in a later stage of learning.
Although DIP is an excellent method that can effectively restore images, in the case of denoising, it does not explicitly minimize the error as a noise component. N2V outperforms DIP for denoising tasks; however, its performance is limited when local smoothness is severely violated.
In our approach, on the other hand, noise removal is attained by explicitly minimizing noise.

\section{Experiments}
%%%%%
%\subsubsection*{Setting}
We collect a dataset of HSIs from web sites, which are commonly used for research purposes. The images have 100 to 225 spectral channels.

We used four separable layers in the network. Each layer had $100$ depth-wise kernels with a size of $(3\times 3\times 1)$, $400$ point-wise kernels with a size of $(1\times 1\times 100)$, batch normalization, and ReLU.
For training, data were augmented by rotating and flipping the input data and were then divided into $20\times 20 \times \{\text{number of bands}\}$ blocks.
We used a minibatch size of 32, and the network was trained using the Adam optimizer \cite{Adam}. We started with a learning rate of 0.01, which was halved every dozens of epochs.

\subsection{Gaussian noise removal}
%%%%%
%%%%%
In our experiments, we added Gaussian noise with a standard deviation of $\sigma=[0.05,0.1,0.15,0.2,0.25]$ and evaluated the performance by averaging the PSNR for all the bands. We compared our method with the conventional model-based methods, namely,  BM4D \cite{BM4D} and FastHyDe \cite{FastHyDe}, and self-supervised DNN methods, that is, DIP \cite{DIP} and N2V \cite{N2V}. We used dense networks for DIP and N2V with the same number of layers shown in their papers, and in each layer we appropriately extended the number of channels.
With our method, we take a noisy image and estimate the standard deviation of the noise from the input using the technique described in \cite{estSigma}. Using the estimated standard deviation, we further added Gaussian noise and used it as a target image, as described in Section \ref{sec:Gauss}. 

Table \ref{tab:Gauss} lists the average PSNR values for some images, from which we can confirm the superiority of the proposed method at high noise levels. When the noise level is low, FastHyDe significantly performs well. Our method is inferior for images with many very dark channels such as 'Salinas'.
Focusing on the DNN-based method, the performance of DIP is inferior to the optimization-based method, but N2V shows the same performance as the optimization-based method at high noise levels. 
%%%%%%%%%%%%%%%%%%%%
%%%%%%%%%% Table supervised
%%%%%%%%%%%%%%%%%%%%
\begin{table}[h]
\caption{Comparison with Supervised Learning for Gaussian noise with $\sigma = 0.2$ (Spv stands for Supervised learning)}

\begin{center}
\begin{tabular}{ccccccc} \bhline{1pt}
%\multicol{2}{Methods} & \multicolumn{5}{c|}{$\sigma$} \\ \cline{2-6}
 Images&  \multicolumn{2}{c}{PaviaC} & \multicolumn{2}{c}{PaviaU} &  \multicolumn{2}{c}{Frisco} \\ \hline 
 $\sigma$& 0.1 & 0.2 & 0.1 & 0.2 & 0.1 &0.2\\ \hline
Spv & 38.07 & 34.97 & 38.55 & 35.43&39.13&35.81 \\% \hline
Ours & 37.60 & 34.87 & 37.65 & 34.81&43.60&38.72\\ \bhline{1pt}
\end{tabular}
\end{center}
\label{tab:supervised}
\end{table}

%%%%%%%%%%%%%%%%%%%%%%
\subsubsection*{Comparison with supervised learning}
%%%%%%%%%%%%%%%%%%%%%%
To demonstrate the efficiency of our self-supervised approach, we compared it with a supervised training method. For the supervised training, we used ten residual blocks with skip connections, where a single block contained a separable convolutional layer, batch normalization, and ReLU, while the kernel size was set such that the number of output channels in the middle layer was 500. 

Assuming that contained noise follows Gaussian distribution with a fixed standard deviation, we created training data sets by adding noise with the fixed standard deviation.
Training was performed using 20 hyperspectral images, which were divided into $40\times 40\times 50$ images after performing data augmentation using reversal or rotation. Using the Adam optimization method, we performed training with 40 epochs at a learning rate of $1e-3$, followed by ten epochs at a learning rate of $1e-4$.

Table \ref{tab:supervised} shows the comparison with the supervised method for several images. As the supervised method is non-blind, that is, we assume that the standard deviation is known, the results are better than our methods for some images. However the method is essentially influenced by training data sets, for the images with characteristics different from the one of the training data, the denoising performance is significantly degraded, while our method is inherently free from the problem and stably yield satisfactory results.
%%%%%
\subsection{Mixed anomaly removal}
\begin{figure}[h]
\centering
\includegraphics[width=0.8\linewidth]{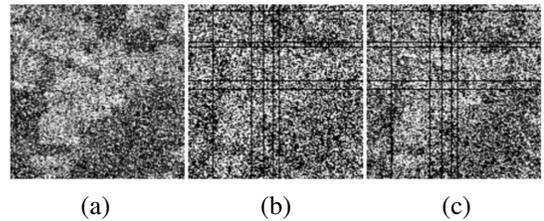}\\
(a)\hspace{2cm}(b)\hspace{2cm}(c)
\caption{Examples of corrupted images in mixed noise removal}
\label{fig:mix}
\end{figure}

%%%%%
We conducted experiments of mixed anomaly removal, in which we simulated situations where several types of noise and anomalies occurred simultaneously, as follows:
\begin{enumerate}
\renewcommand{\labelenumi}{(\alph{enumi})}
\item{Gaussian noise + impulsive noise:}
The images were corrupted by Gaussian noise for which $\sigma=0.05$ and by impulsive noise with a density of $d=20\%$.
\item{Gaussian noise + line deficit: }
In addition to the Gaussian noise for which $\sigma=0.15$, we selected vertical and horizontal lines in some bands and deleted the pixels in the lines.
\item{Gaussian noise + impulsive noise + line deficit: }
We added all three anomalies, that is, the Gaussian noise for which $\sigma=0.1$, impulsive noise with an intensity of $d=10\%$, and the line deficit.
\end{enumerate}
The example of corrupted images in the experiments is shown in Figure \ref{fig:mix}.

%%%%%%%%%%%
%%%% Figure Mixed noises
%%%%%%

%%%%%%%% Table: mixed anomalies
\begin{table}[ht]
\caption{Results for mixed anomaly removal}
\begin{center}
\begin{tabular}{cccc} \bhline{1pt}
Methods  &  (a) & (b) &  (c) \\ \hline \hline
%ASTV \cite{ASTV} & 29.05 & 28.58 & 28.59 \\
SSTV\cite{ASTV}  & 32.14 & 29.19 & 30.92 \\
HSSTV \cite{HSSTV} & 32.23 & 29.70 & 31.20 \\
%BM4D & 22.52 & 29.17 & 25.17 \\ 
LRTDTV \cite{LRTDTV} &38.12&34.10&35.22\\ \hline
DIP \cite{DIP}&  22.20 & 29.25  & 25.77 \\
N2V \cite{N2V} & 22.30 & 33.59 & 27.25  \\ \hline
Ours & {\bf 39.29} & {\bf 35.72} & {\bf 35.67} \\ \bhline{1pt}
\end{tabular}
\end{center}
\label{tab:result_mix}
\end{table}
%%%%%%%%

We tested our method for images that are different from those used for the Gaussian denoising, and compared our method with ASTV \cite{ASTV}, SSTV \cite{SSTV}, HSSTV \cite{HSSTV}, and LRTDTV \cite{LRTDTV} as well as DIP and N2V.
Note that we do not list BM4D and FastHyDe because these are optimally designed for Gaussian noises and does not work well for the mixed noises removal.
Table \ref{tab:result_mix} shows the average PSNR of the five images, as well as that for the conventional methods. Our method clearly attains the best results for all the experimental settings.

%%%%%%%%%%%%%%%%%%%%
%%% Figure real noise
%%%%%%%%%%%%%%%%%%%%
\begin{figure*}[!th]
\centering
\tabcolsep = 0.2mm
\begin{tabular}{cccc}
 \includegraphics[width=.18\linewidth]{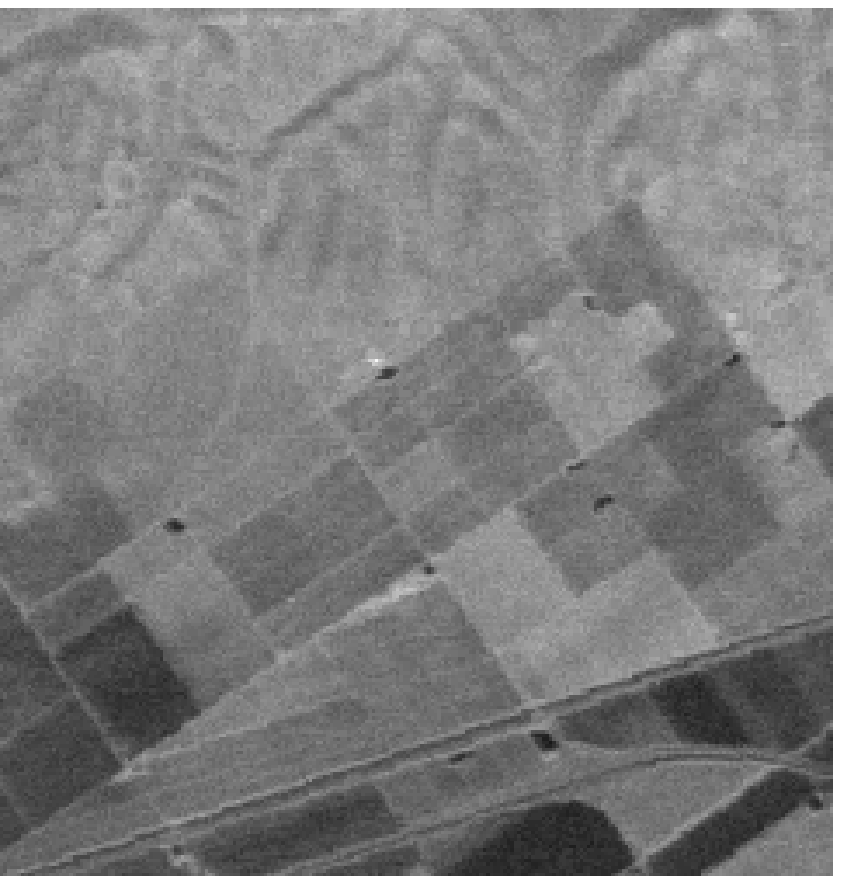} \hspace{0.3cm}&
\hspace{0.3cm} \includegraphics[width=.18\linewidth]{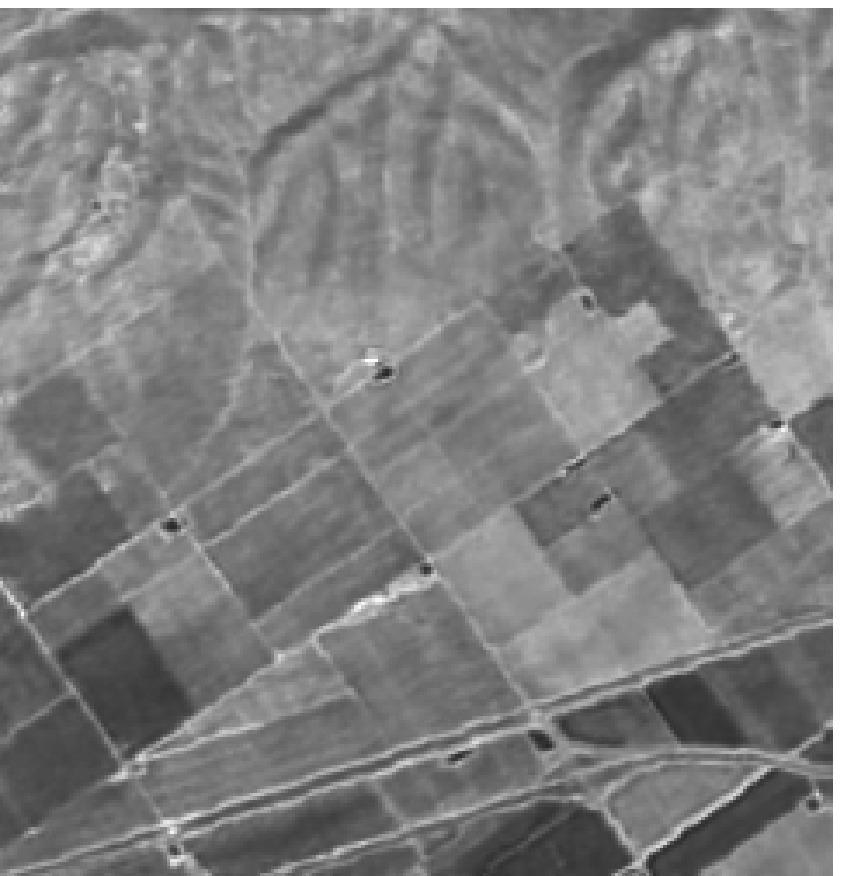} \hspace{0.3cm}&
\hspace{0.3cm} \includegraphics[width=.18\linewidth]{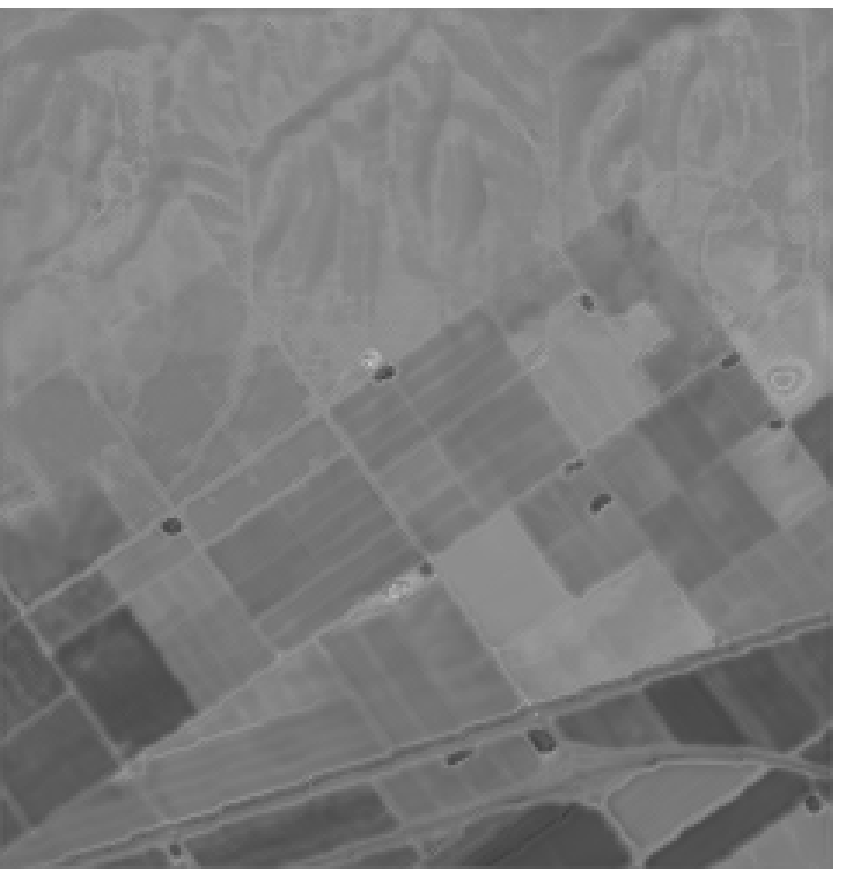} \hspace{0.3cm}&
\hspace{0.3cm} \includegraphics[width=.18\linewidth]{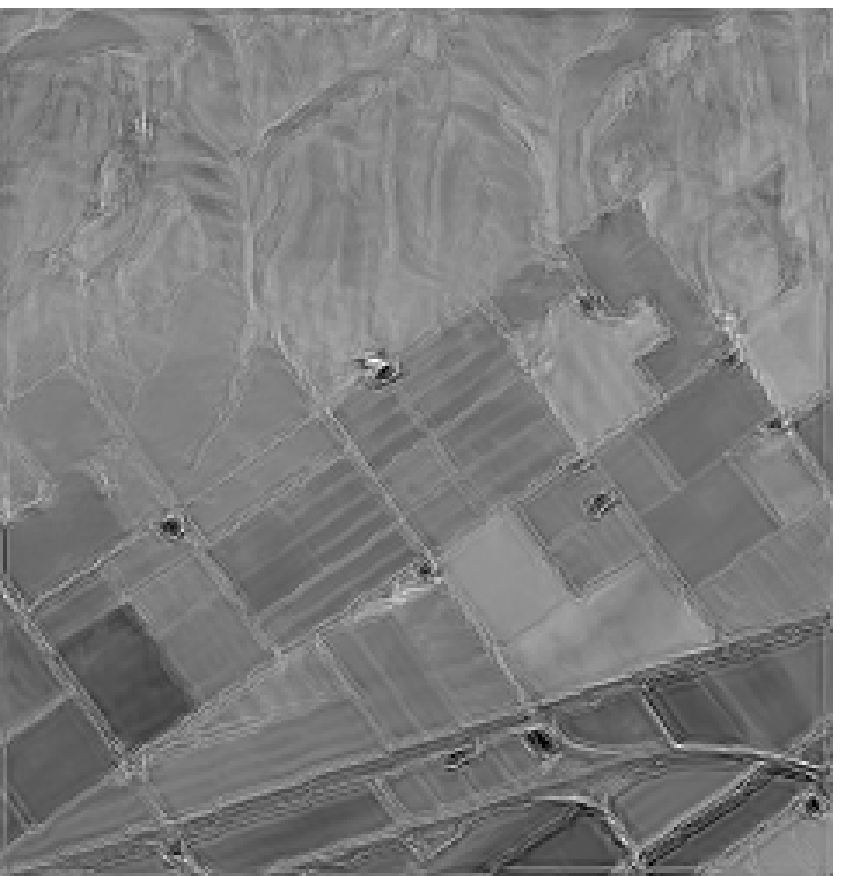}\\
\includegraphics[width=.18\linewidth]{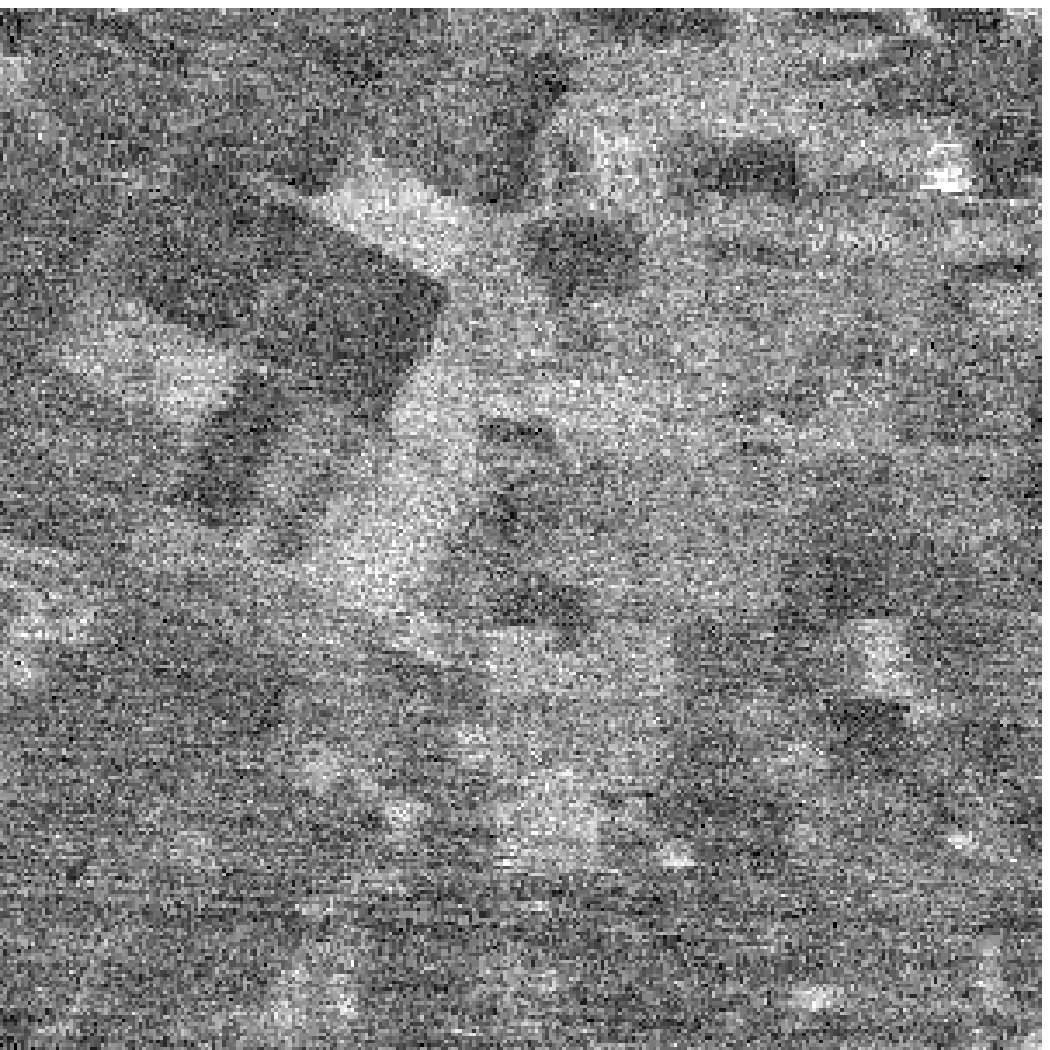} \hspace{0.3cm}&
\hspace{0.3cm} \includegraphics[width=.18\linewidth]{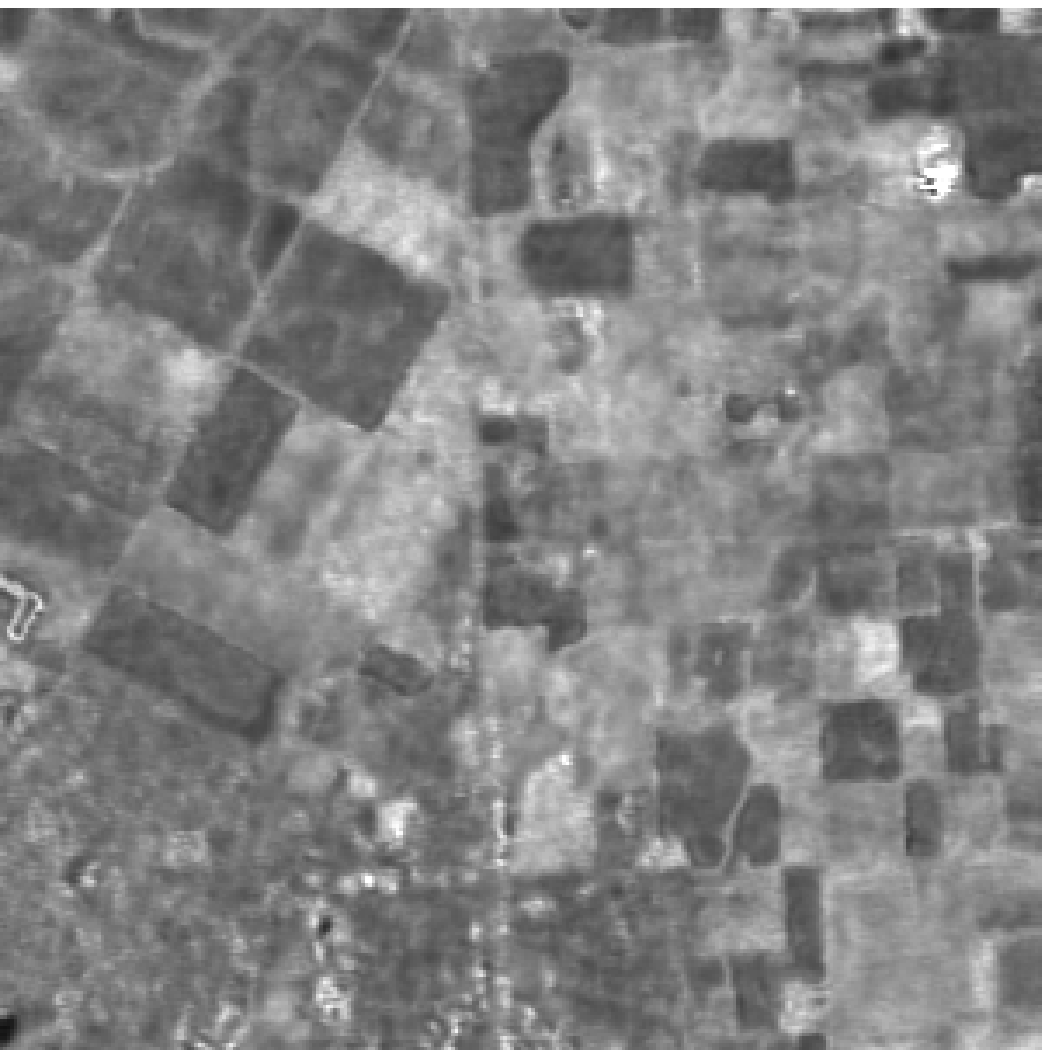} \hspace{0.3cm}&
\hspace{0.3cm} \includegraphics[width=.18\linewidth]{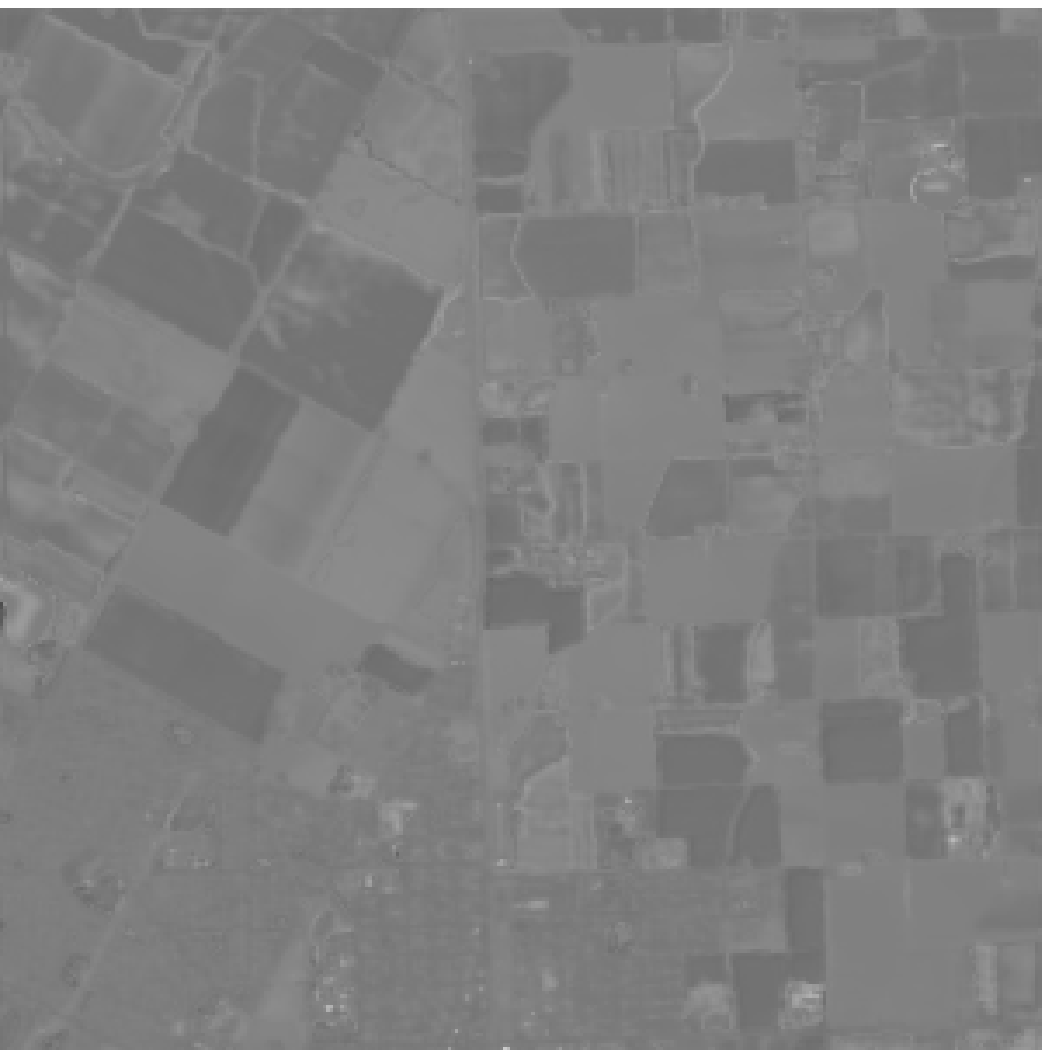} \hspace{0.3cm}&
\hspace{0.3cm} \includegraphics[width=.18\linewidth]{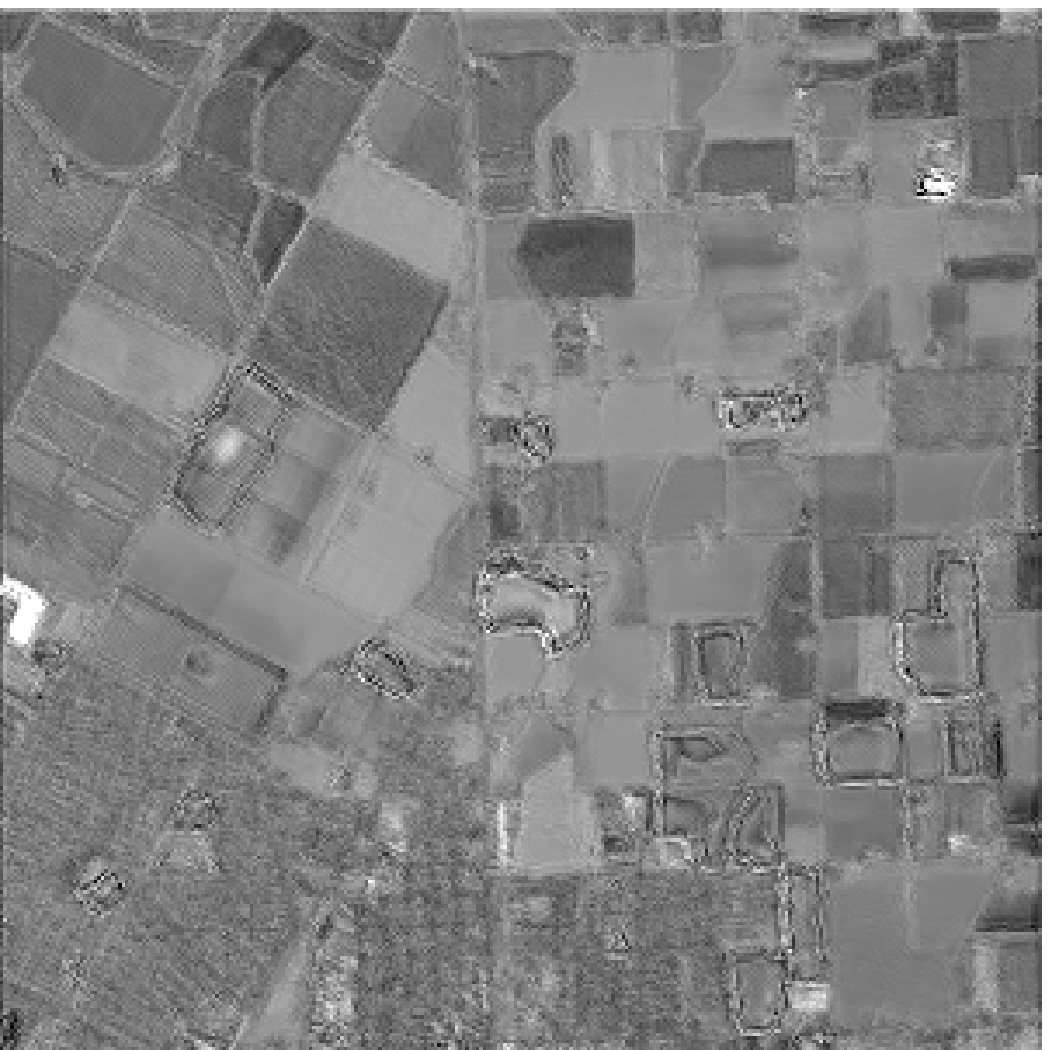}\\
\includegraphics[width=.18\linewidth]{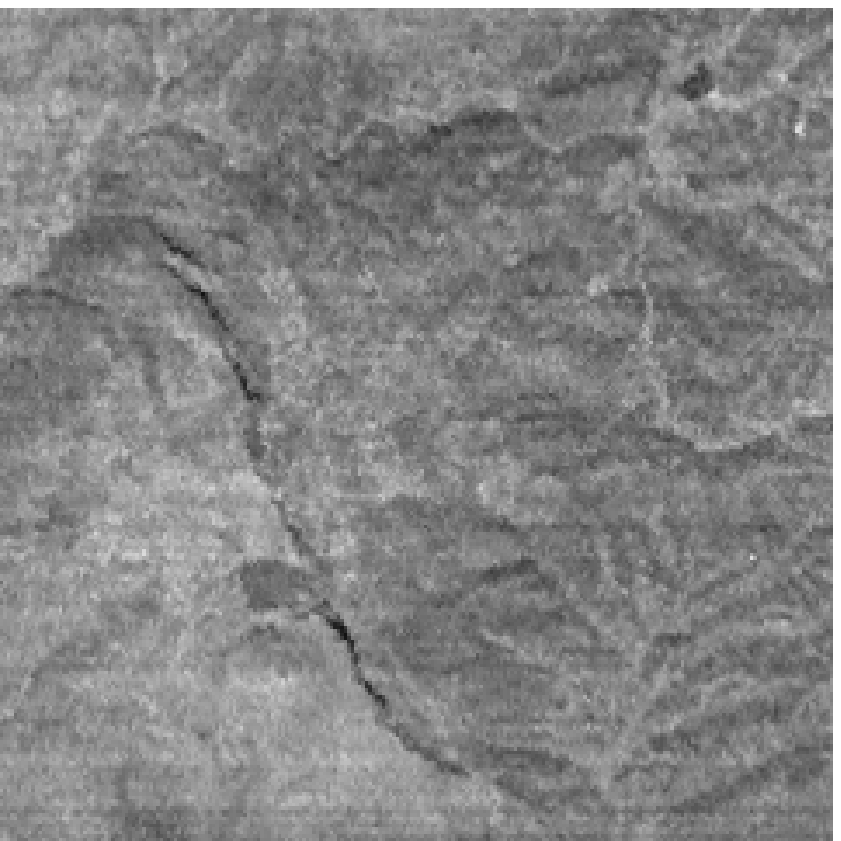} \hspace{0.3cm}&
\hspace{0.3cm} \includegraphics[width=.18\linewidth]{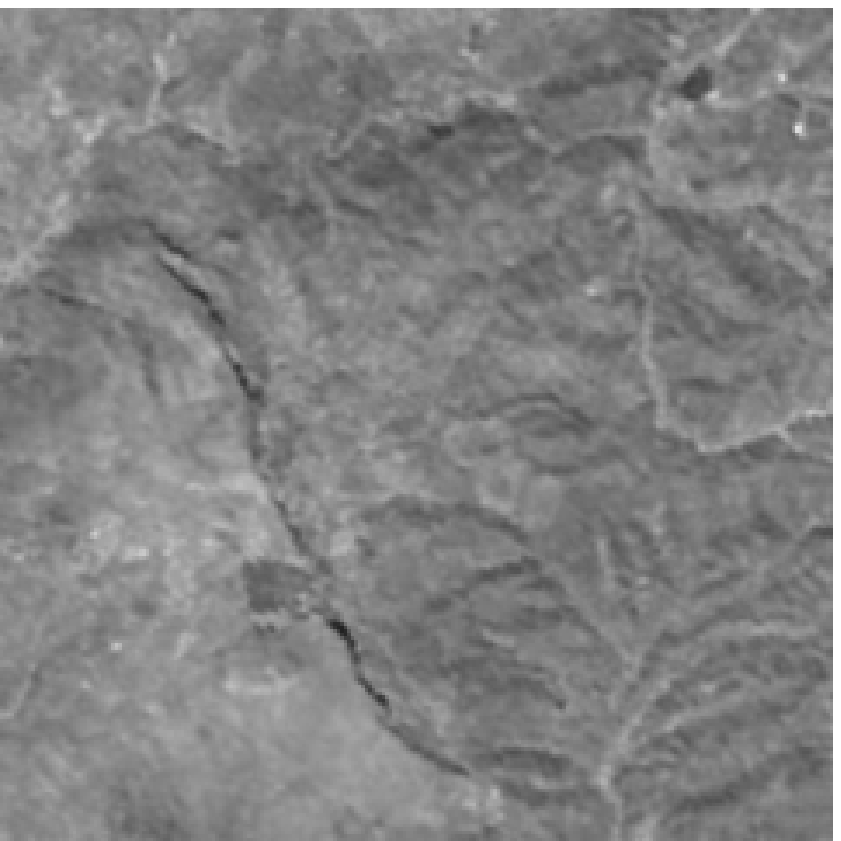} \hspace{0.3cm}&
\hspace{0.3cm} \includegraphics[width=.18\linewidth]{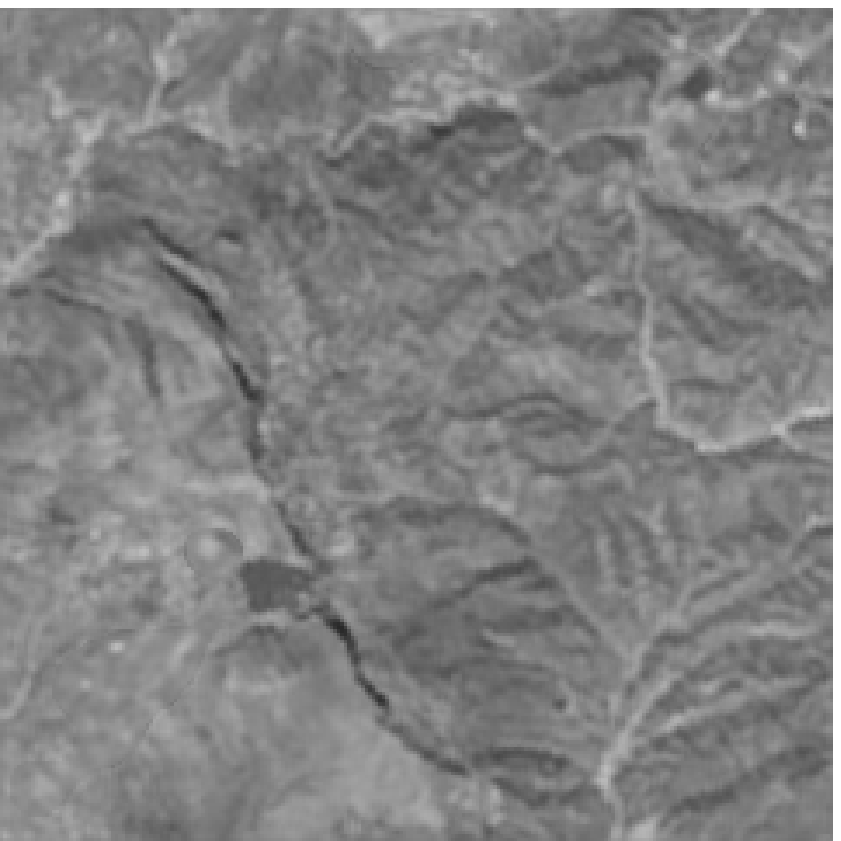} \hspace{0.3cm}&
\hspace{0.3cm} \includegraphics[width=.18\linewidth]{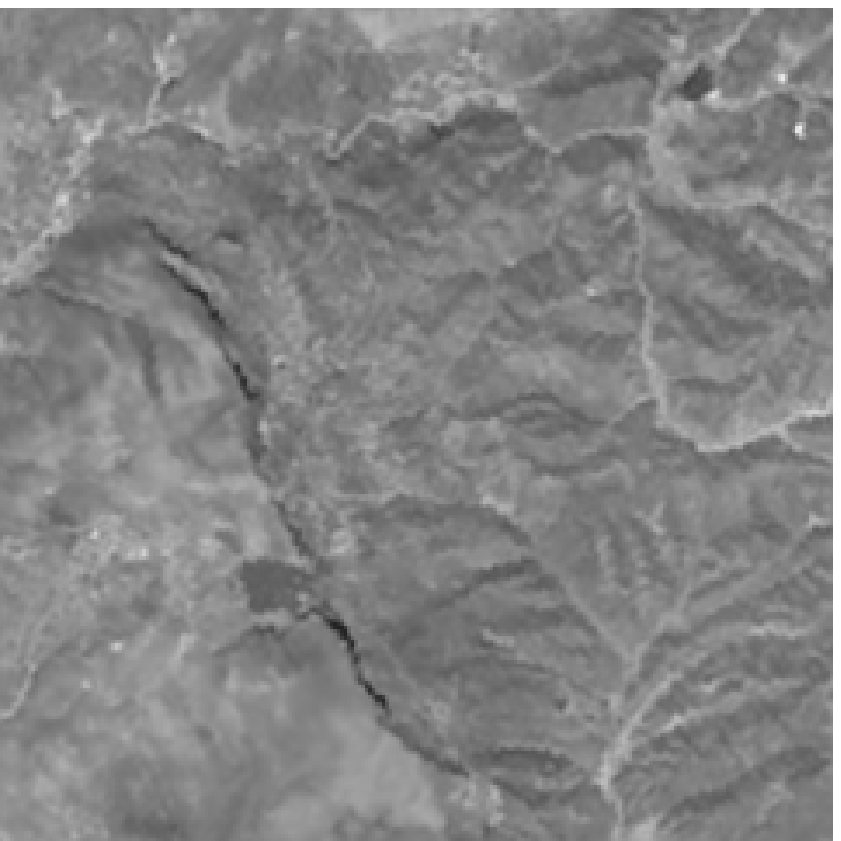}\\
\includegraphics[width=.18\linewidth]{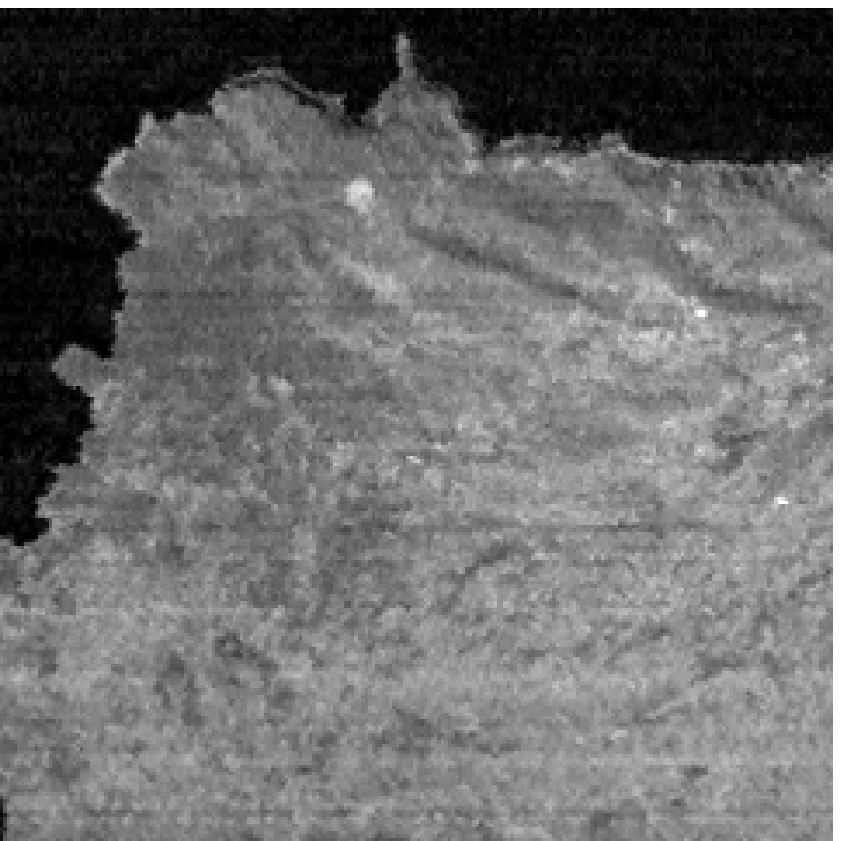} \hspace{0.3cm}&
\hspace{0.3cm} \includegraphics[width=.18\linewidth]{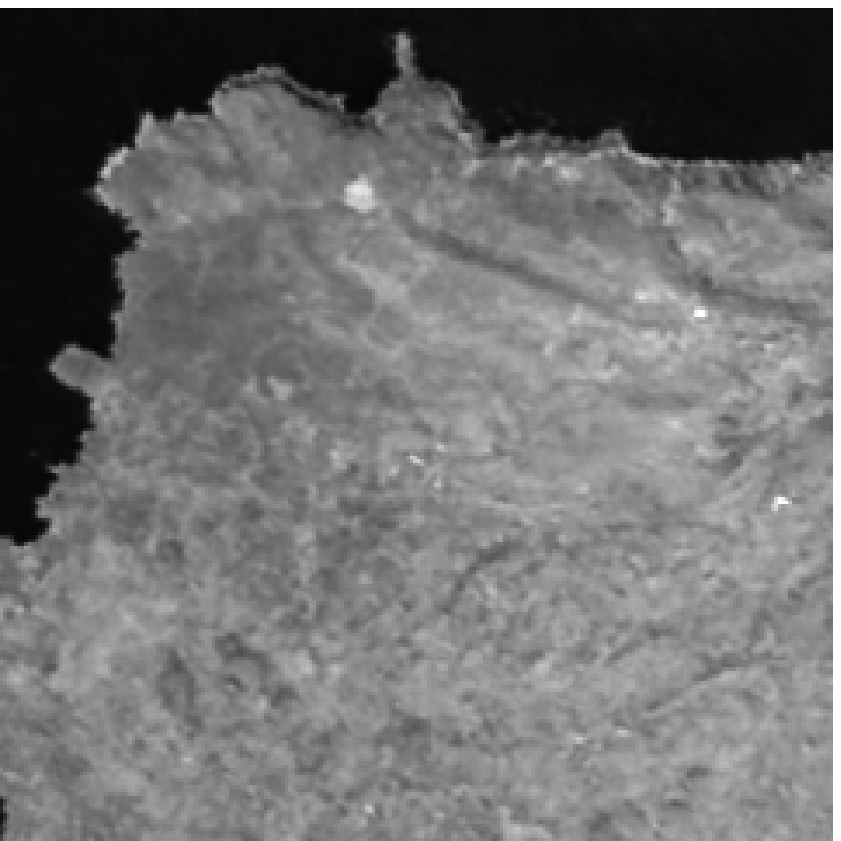} \hspace{0.3cm}&
\hspace{0.3cm} \includegraphics[width=.18\linewidth]{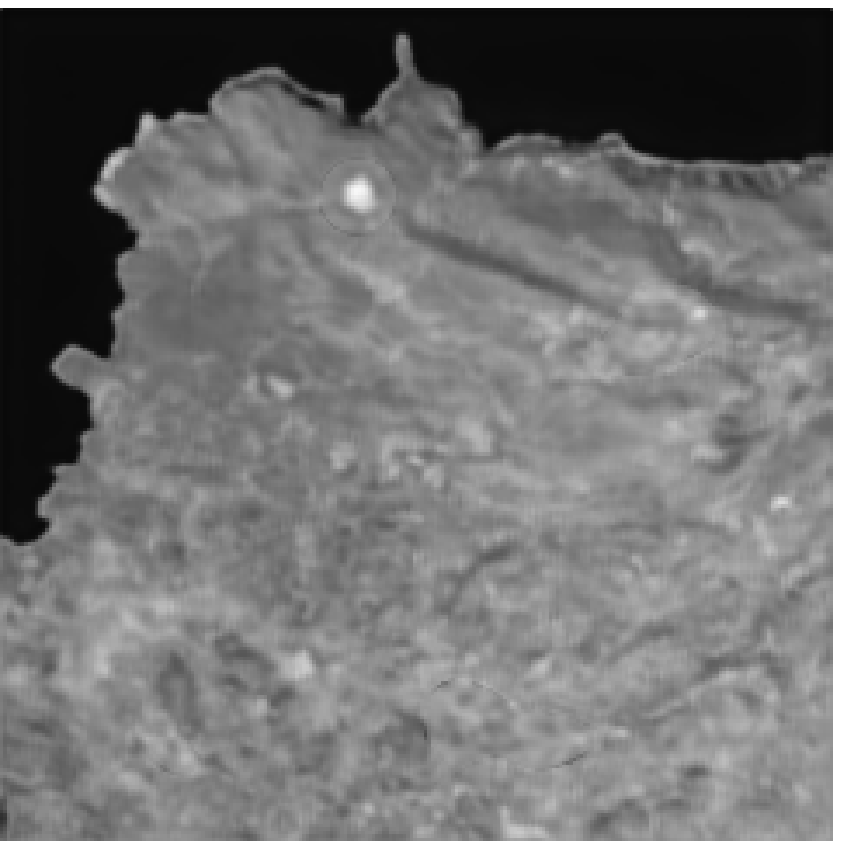} \hspace{0.3cm}&
\hspace{0.3cm} \includegraphics[width=.18\linewidth]{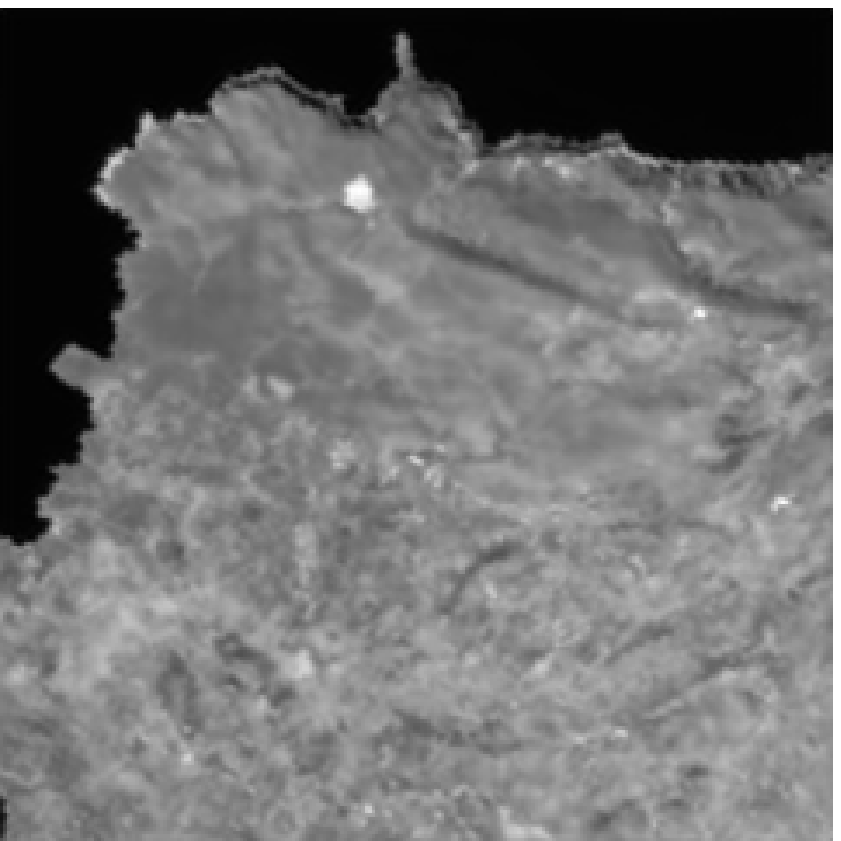}\\
Original  & DIP & N2V & Ours\\
\end{tabular}
\caption{Real Noise Removal}
\label{fig:real}
\end{figure*}
%%%%%%%%%%

%%%%%%%%%%%
\subsection{Real noises}
%%%%%%%% Figures of resultant images
Ground truth images are often unavailable, and the characteristic of noise is sometimes unknown in actual situations. It is desirable that denoisers are robust to noise characteristics. DIP and N2V have a significant advantage that they work well for blind denoising problems, that is they do not require any knowledge about noises.

To show that our method also works in the blind situation, we conducted the experiment of real noise removal using some HSIs which have a few noisy bands, and subjectively compared with DIP and N2V. 
Figure.\ref{fig:real} shows some examples for the real noise removal. According to our experiments, the three methods have some drawbacks; for example, DIP sometimes yields over-smoothing artifacts, N2V tends to low contrast images when an input is severely degraded, while our method sometimes yields too sharp contrasts.
However, it can be seen from the figure that our method successfully restore a sharp image. The results of DIP and N2V also show their abilities to remove real noises. 
%%%%%
\section{Conclusion}
We have introduced the new self-supervised learning for HSI restoration. The strength of our method is owing to both of the use of the separable convolutional layers as well as the self-supervised strategy. The success is also owing to rich information of an HSI with many channels. Our approach is not very efficient for grayscale/color images. The effectiveness of the proposed method is also limited when applying to other 3D images such as videos, MRI and  CT because the separable CNN tends to strongly induce low-rankness w.r.t. the channel direction and often yield over-smoothing effects.

Our method robustly works well for many types of noises. We have investigated the effectiveness of our approach through some restoration tasks including real noise removal. To the best of our knowledge, this is the first work that shows the superiority of deep learning for HSI restoration over the conventional model-based methods.
%The codes are available at [to be filled in].
%\subsubsection*{Acknowledgement}
%To be filled in.
%
%\vspace{.5cm}

%%%%

{\small
\bibliographystyle{ieee}
%\bibliography{ref_RSJ}
\bibliography{ref}
}

\end{document}